\documentclass[twocolumn,prl,showpacs,superscriptaddress,preprintnumbers]{revtex4} 
\usepackage{amsmath,amssymb,epsfig,color} 

\begin{document} 

\title{Raman study of lattice dynamics in quasicrystals.} 
\author{Yu. S. Ponosov} 
\affiliation{Institute of Metal Physics UD RAS, 620041, S. Kovalevskaya str. 18, Ekaterinburg, Russia} 
\author{N. I. Shchegolikhina} 
\affiliation{Institute of Metal Physics UD RAS, 620041, S. Kovalevskaya str. 18, Ekaterinburg, Russia} 
\author{A. F. Prekul} 
\affiliation{Institute of Metal Physics UD RAS, 620041, S. Kovalevskaya str. 18, Ekaterinburg, Russia}

\begin{abstract} 
We present the first Raman investigation of icosahedral quasicrystals. Broad structured bands in the energy range up to $\sim$ 500 $cm^{-1}$ have been observed in a series of AlFeCu, AlPdMn and AlPdRe systems. Original information on the vibrational density of states g($\omega$) was   obtained for AlPdRe; for AlFeCu and AlPdMn estimated g($\omega$) shows a good agreement with the previous  neutron results,  but demonstrates finer structure.  Strong increase in the parameter of electron-vibrational coupling  for the low-energy vibrations and its correlation with changes in electronic conductivity  has been observed in the series  from AlFeCu to AlPdRe. This suggests  the  increase of the degree of localization for these vibrational excitations and involved electronic states.  
\end{abstract}    

\maketitle 
Quasicrystalline structures are often presented as an intermediate state between a periodic 
and a random medium because they combine sharp Bragg peaks in diffraction patterns with 
the absence of lattice translation periodicity~\cite{sheh}. However, it is still not clear how such an unique 
structure determines their unusual electron transport properties. Although investigations 
of quasicrystals have been carried out for more than two decades, experimental information 
on their electron structure and vibrational spectra is still rather scarce. The problem 
of existence of extended/localized electronic states and vibrational excitations and their 
coupling seems to be very important for understanding the stability and unique conduction properties of these systems~\cite{janot,trambly}. 

Neutron spectroscopy so far has been the most useful technique to study phonon dispersion 
relations and vibrational spectra in quasicrystals. Triple-axis experiments have shown that 
rather sharp acoustic phonon peaks exist at very small wave vectors \textbf{q} similar to the propagative 
modes in the lattice periodic systems ~\cite{bois,boud}. For larger q values the observed excitations  strongly 
broaden both in energy and in momentum space, showing spectra resembling the phonon 
density of states. A generalized vibrational density of states (GVDOS) may be obtained from 
the time-of-flight neutron experiment. The measured GVDOS of quasicrystals usually has a broad 
profile with poor structure~\cite{Suck}. Since the partial DOS of different metals are weighted in 
GVDOS by specific neutron coefficients, quantitative information on the total VDOS 
g(E) of a quasicrystal may be extracted using complex procedures, for example, the method 
of isotopic contrast~\cite{brand,par}. The element-partial 
g(E) in icosahedral i-$Al_{62}Cu_{25.5}Fe_{12.5}$ was also studied by inelastic nuclear resonant absorption (INA) of synchrotron radiation~\cite{ina}. 

The data on sound velocities and surface phonons  in quasicrystals have been obtained  from ultrasonic and Brillouin  experiments~\cite{ultra,br}. Vibrational excitations  have also been observed in the far-infrared region of the optical spectra of AlPdRe ($\sim$200$cm^{-1}$)~\cite{bian,bas}, AlCuFe ($\sim$ 245 $cm^{-1}$)~\cite{tim}, AlPdMn ($\sim$ 270 $cm^{-1}$)~\cite{optics}.  

It is surprising that no Raman study of quasicrystals has been reported after unsuccessful attempt~\cite{br}. Raman spectroscopy is well known to be a powerful technique for studying different elementary excitations in solids; however, it has  found rather limited application for metals because of a small penetration depth of light in an opaque material.  Authors of~\cite{br} have explained their failure to observe the Raman spectra from quasicrystals by their high  conductivity at optical frequencies. However, sometimes, high interband absorption of metal may provide the resonance conditions (at the energies of an incident/scattered light) for light scattering, which leads to enhancing spectral intensity~\cite{Ti,Os}. Since optical properties of quasicrystal in the visible range  are similar to those of metals, one can try to search for the Raman spectra more carefully. In comparison with the neutron technique Raman method possesses much better resolution, needs very small material quantities for analysis, and may be used to study systems consisting of elements not suitable (for example, rhenium) for neutron experiment. In principle, Raman spectroscopy in metals may give the information both on the  lattice dynamics and electronic excitations~\cite{Os,Pres}. A Raman spectrum in a perfect crystal   consists of a number of narrow lines in accordance with the selection rules for the given crystal structure.  Frequencies of these lines correspond to the energies of long-wavelength optical phonons. The vibrations from the whole Brillouin zone (pseudozone) may become active in the Raman scattering upon breakdown of translation periodicity . Well-known examples for that are the Raman measurements of the lattice dynamics in amorphous solids~\cite{amob}. They yield information both on the vibrational DOS and local structural order.  

In this work we present the first  Raman study of three quasicrystalline systems AlCuFe, AlPdMn and AlPdRe. We observed broad spectra with several superimposed features in the range up to 500 $cm^{-1}$ for each system.  The comparison of the reduced spectra with published neutron and theoretical results shows overall profile coincidence. The intensities of the low-frequency features in  the Raman spectra exhibit a strong increase in the series from AlCuFe to AlPdRe. This indicates  the changes in the coupling coefficient of light with vibrations owing to corresponding changes in the correlation length of the low-frequency vibrational excitations. 

The study was performed using homogeneous samples of icosahedral phase with the nominal compositions $Al_{63}Cu_{25}Fe_{12}$, $Al_{62}Cu_{25.5}Fe_{12.5}$, $Al_{70.2}Pd_{21.4}Mn_{8.4}$, $Al_{70}Pd_{21.5}Re_{8.5}$ and $Al_{70}Pd_{20}Re_{10}$, prepared from components with a purity not lower than 99.9$\%$.  The alloys were melt in an arc furnace under high purity argon, quenched on the water-cooled hearth  by the 'hammer-and-anvil' method, and annealed at temperatures of 1000 (AlCuFe), 1050 (AlPdMn) and 1170 K (AlPdRe), respectively, for 12 hours~\cite{prekul}. The structure of the alloys was studied by scanning electron microscopy, X-ray diffraction (STADI-P), and transmission electron microscopy (JEM 200-CX). No secondary phases were detected. High quality of the samples is confirmed also by the resistivity (Fig.1) and the magnetic susceptibility measurements.   

The Raman spectra 
were excited by a low-power (up to 5 mW) laser radiation  at wavelengths 
of 514 and 633 nm, and they were recorded by Renishaw  microscope spectrometer, 
providing a focal spot on the samples of $\sim$ 2 $\mu$m diameter. The 
spectral resolution was about 3 $cm^{-1}$.  Little difference was found between spectra  taken from the mechanically polished samples  and  from their small cleaved surfaces. The low-frequency measurements have been limited  to $\sim$ 70 $cm^{-1}$. All studies have been taken at room temperature. 

\begin{figure}[t] 
\includegraphics[width=0.4\textwidth]{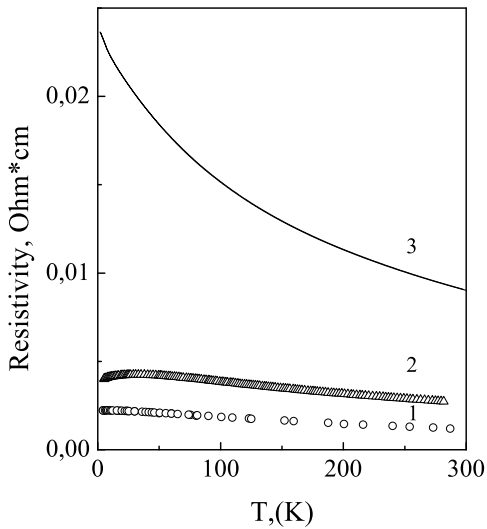} 
\caption{\label{label} Temperature dependence of electrical resistivity  $\rho$(T) of $Al_{63}Cu_{25}Fe_{12}$ (1), $Al_{70.2}Pd_{21.4}Mn_{8.4}$ (2) and $Al_{70}Pd_{20}Re_{10}$ (3). } 
\end{figure} 

The Raman spectra measured from the cleaved surfaces of two AlPdRe samples with different compositions are shown in Fig.2. They have similar profiles composed from broad split bands near 250 and 400 $cm^{-1}$ and prominent feature near 115 $cm^{-1}$. The same spectra were observed with both laser excitations, which evidences inelastic light scattering. Similar spectra were also measured in the parallel and perpendicular polarization geometries (Fig.5). We  found no information on  the vibrational spectra of AlPdRe system in literature except for the  far-infrared data~\cite{bian,bas} where the feature near 200 $cm^{-1}$ was observed. The measurements were extended to the high-energy range to 3000 $cm^{-1}$ where a broad weak band was stably observed near 1800 $cm^{-1}$ with the 514 nm excitation. The origin of this band is so far unclear; its high energy suggests that some electronic excitations  may be resonantly enhanced with the 514 nm incident wavelength, if it is not a luminescence band. 

\begin{figure}[b] 
\includegraphics[width=0.4\textwidth]{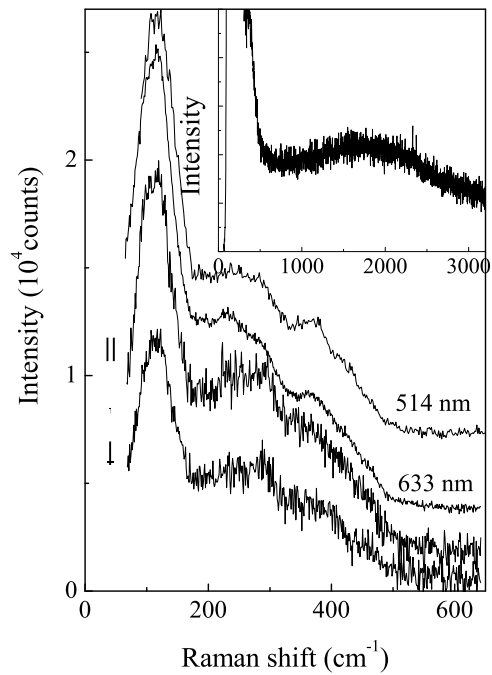} 
\caption{\label{label} Raman spectra of $Al_{70}Pd_{20}Re_{10}$ measured in the parallel polarization geometry with two excitation wavelengths (two top spectra). Two bottom spectra were measured in different polarization geometries from the $Al_{70}Pd_{21.5}Re_{8.5}$ sample. Background was subtracted. Inset shows the high-energy spectral range. } 
\end{figure} 
The Raman spectra for all the investigated quasicrystalline systems are shown in Fig.3. They consist of three main bands, and  high-energy boundary of all spectra lies near 500 $cm^{-1}$. Two high energy peaks near 250 and 400 $cm^{-1}$ are split and the energies of the spectral features slightly vary for the different systems. Low frequency peak is also split in $Al_{70.2}Pd_{21.6}Mn_{8.4}$. The curve for this sample can be directly compared to the  neutron spectrum of Suck~\cite{Suck} (inset in Fig.3) though both spectra should be corrected to obtain the total VDOS. One can see a rather good agreement in the shape of both spectra but the Raman spectrum is more structured. The energies of two low-frequency peaks at about 95 and 125 $cm^{-1}$ are close to the energies of dispersionless "optic" excitations near 3 and 4 THz, found in triple-axis neutron experiment~\cite{bois,boud}. As for high-energy peaks, they may be due to a long-range ordered structure of quasicrystals, which leads to a spiky structure in the density of states in some icosahedral models.  
\begin{figure}[b] 
\includegraphics[width=0.4\textwidth]{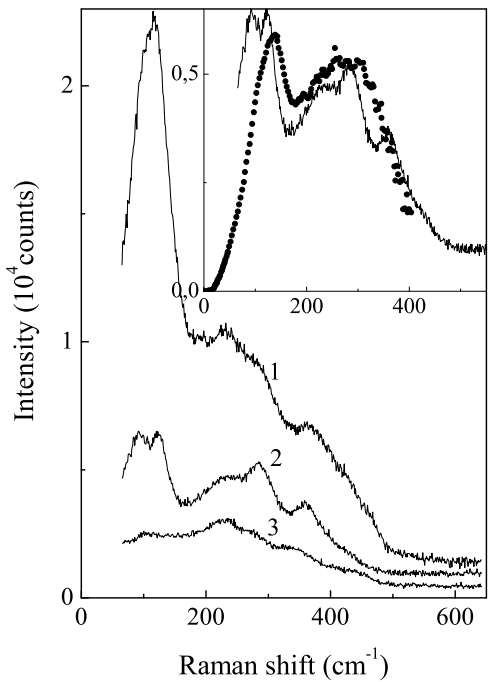} 
\caption{\label{label} Raman spectra of $Al_{70}Pd_{20}Re_{10}$ (1), $Al_{70.2}Pd_{21.4}Mn_{8.4}$(2) and  $Al_{62}Cu_{25.5}Fe_{12.5}$ (3) at T=300K, measured with the 633 nm excitation . Background was subtracted.  Inset: Raman spectra \textit{vs} GVDOS from the neutron experiment~\cite{Suck} for $Al_{70.2}Pd_{21.4}Mn_{8.4}$(filled circles). } 
\end{figure}  
\begin{figure}[b] 
\includegraphics[width=0.4\textwidth]{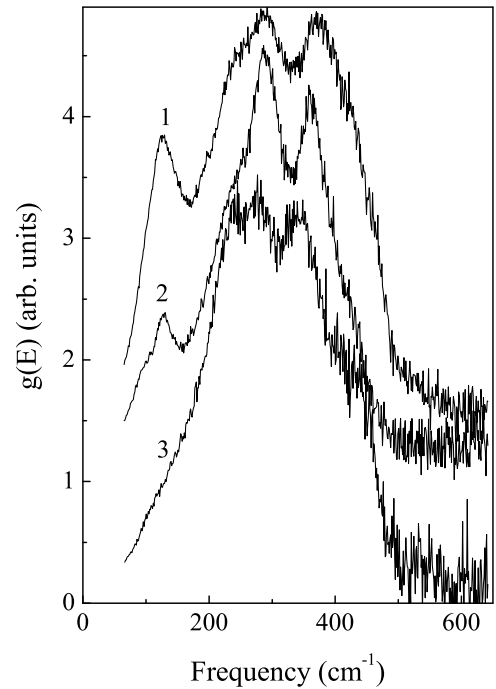} 
\caption{\label{label} Vibrational densities of states for $Al_{70}Pd_{20}Re_{10}$ (1), $Al_{70.2}Pd_{21.4}Mn_{8.4}$ (2) and $Al_{62}Cu_{25.5}Fe_{12.5}$ (3) calculated from the Raman spectra using Raman coupling parameter C($\omega$)=1 . } 
\end{figure}  
The spectral intensities  show an essential increase in the series from AlCuFe to AlPdRe (Fig.3).  This intensity growth  at the energy range above 250 $cm^{-1}$ may be well explained by the changes of the optical constants~\cite{nom} which determine the variation in both the reflection/absorption of the incident/scattered light and  scattering volume.  As for intensity increase of the low-frequency excitations near 100$-$120 $cm^{-1}$, actually, it has another nature and is due to a strong variation of the Raman matrix element. 
  
 The total Raman spectra profiles and their comparison with the neutron results suggest that the Raman spectra of quasicrystals provide an information on the  vibrational density of states. In the case of amorphous or disordered solids, the Raman intensity is described by~\cite{ram} 
\begin{eqnarray} 
I(\omega)\propto C(\omega)[n(\omega,T)+1]\omega^{-1}g(\omega) 
\end{eqnarray} 
where $\omega$ is the vibration frequency, n($\omega$,T) is the Bose factor, C($\omega$) is the coupling coefficient of light with vibrations, and g($\omega$) is the vibrational density of states. In Fig.4 we show g($\omega$), obtained from the measured spectra using Eq.1 with C($\omega$)=1, for all the quasicrystals studied.   Unfortunately, there is no possibility to compare the Raman results with the data on VDOS in AlPdMn and AlPdRe from the neutron measurements, since they are not available (only experimental spectrum for $Al_{70.2}Pd_{21.4}Mn_{8.4}$ was published~\cite{Suck}).  

For the AlCuFe quasicrystal, g($\omega$) from the Raman experiment is  drawn in Fig.5 together with g($\omega$) from the neutron data (the latter is in a satisfactory agreement with the calculated spectrum~\cite{maz}). One can see a strong difference between the Raman and neutron g($\omega$): the low-energy range (approximately up to 250 $cm^{-1}$) of the Raman g($\omega$) is strongly suppressed  in comparison with the neutron g($\omega$). This means that the Raman coupling parameter has a strong  frequency dependence in this range. Comparing two g($\omega$) from Fig.5,  we obtained a rough estimate of C($\omega$). It shows that the coupling parameter essentially increases in the range from 70 to $\sim$ 250 $cm^{-1}$ (inset in Fig.5).  

\begin{figure}[t] 
\includegraphics[width=0.4\textwidth]{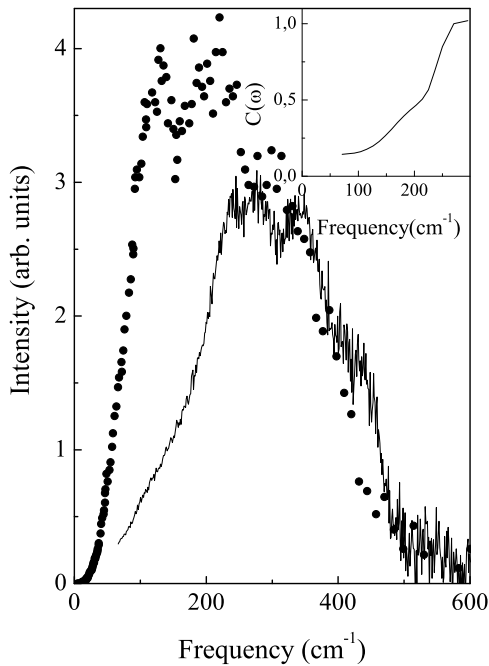} 
\caption{\label{label} g($\omega$) for  $Al_{62}Cu_{25.5}Fe_{12.5}$ calculated from the Raman spectra with C($\omega$)= 1  and total VDOS for $Al_{62}Cu_{25.5}Fe_{12.5}$ from the neutron experiment~\cite{par} (filled circles). Inset: frequency dependence of the coupling parameter. } 
\end{figure}  

Such situation  takes place in the case of the Raman scattering in  $\alpha$-Si and $\alpha$-Ge~\cite{amSi,amGe}, where the frequency behavior of C($\omega$) is very similar for both amorphous semiconductors when scaling is employed. The frequency dependence of C($\omega$) observed in them  reflects a stronger tendency to localization for the optical vibrations than for the acoustical ones because of a different degree of the violation of the momentum selection rules~\cite{ch}. It was shown that the Raman matrix element for the acoustical phonons is proportional to the square of inverse correlation length of vibrational excitations~\cite{nov}. As Figs. 3, 4 demonstrate, the intensity increase of the low-energy peaks near 120 $cm^{-1}$ was revealed in the series from AlCuFe to AlPdRe. This means that, contrary to  C($\omega$) behavior in the above mentioned semiconductors, an essential variation of the electron-vibrational coupling parameter, and, therefore, different degree of localization of these excitations occurs in the studied systems. 

The neutron experiments using isotopic contrast method~\cite{brand,par} identified that the copper atoms in $Al_{62}Cu_{25.5}Fe_{12.5}$ vibrate near 14 meV ($\sim$ 115 $cm^{-1}$). A very weak feature at  $\sim$ 110 $cm^{-1}$ is observed in the Raman spectra and VDOS of this material (Figs.3-5). The authors of~\cite{brand,par} noted that the copper atoms are bonded weaker, on the average, than the iron atoms, which results in a large difference of their vibrational frequencies (iron vibrations are centered at 28 meV (226 $cm^{-1}$) in spite of rather small mass difference. The replacement of copper and iron by heavier atoms in AlPdMn and AlPdRe leads to a small hardening of the low-frequency peaks. Such behavior contradicts the mass factor tendency, confirming an anomaly observed in the force constants of Cu vibrations in AlCuFe. 

The energy scale of the low-energy excitations ($\sim$ 15-20 meV) is obviously not accidental for the electron structure of quasicrystals. It has been observed in tunneling~\cite{tun} and NMR~\cite{nmr} experiments, when analyzing the temperature behavior  of the conductivity, magnetic susceptibility, Hall effect~\cite{prekul} and, finally, in studies of specific heat as a characteristic energy of the Schottky-type electronic term~\cite{shot}. The decrease of the Raman coupling parameter for the low-energy vibrations in AlCuFe correlates with their small stiffness coefficient and increased mean-square displacements.  This may be interpreted as an evidence that the vibrational excitations in AlCuFe are more extended than similar excitations in AlPdMn and AlPdRe. In addition, the behavior of the parameter of electron-vibrational coupling for the low-frequency excitations correlates with changes in the magnitude of resistivity at least on the system level (Figs. 1, 4). Thus, the increase in the coupling parameter, taken together with the mentioned correlations, implies that the degree of localization for the low-energy vibrations (and, possibly, involved electronic states) increases in the series from AlFeCu to AlPdRe. 

In summary, inelastic light scattering is virtually observable in icosahedral quasicrystals. The 
obtained results proved that the Raman scattering provides a powerful tool for the investigation of 
structural, vibrational and electronic properties of quasiperiodical systems. The method can 
be useful in investigations of the temperature and pressure behavior, structural quality 
of i-quasicrystals, and is not limited to solely icosahedral class of quasicrystalline systems. 

This research was supported by RFBR grant No. 08-02-00437 and UD RAS grant No. 28

\end{document}